\documentclass[sigconf, screen]{acmart}

\usepackage{xspace}
\usepackage[capitalise]{cleveref}
\usepackage[colorinlistoftodos,prependcaption,textsize=tiny]{todonotes}
\usepackage{subcaption}
\usepackage{listings}
\usepackage{microtype}

\definecolor{pblue}{rgb}{0.13,0.13,1}
\definecolor{pgreen}{rgb}{0,0.5,0}
\definecolor{pred}{rgb}{0.9,0,0}
\definecolor{pgrey}{rgb}{0.46,0.45,0.48}
\definecolor{rltred}{rgb}{0.5,0,0}
\definecolor{rltgreen}{rgb}{0,0.5,0}
\definecolor{rltblue}{rgb}{0,0,0.5}
\definecolor{mauve}{rgb}{0.5, 0, 0.5}

\Crefname{lstlisting}{Listing}{Listings}

\newcommand{\summary}[2]{%
	\vspace{-0.2cm}%
	\begin{center}%
		\colorbox{gray!20}{%
			\parbox{0.98\linewidth}{%
				\textbf{\textsf{Summary (\textit{#1})}:}~%
				#2%
			}%
		}%
	\end{center}%
}
		
\copyrightyear{2024}
\acmYear{2024}
\setcopyright{acmlicensed}
\acmConference[Gamify '24]{Proceedings of the 3rd ACM International Workshop on Gamification in Software Development, Verification, and Validation}{September 17, 2024}{Vienna, Austria}
\acmBooktitle{Proceedings of the 3rd ACM International Workshop on Gamification in Software Development, Verification, and Validation (Gamify '24), September 17, 2024, Vienna, Austria}
\acmDOI{10.1145/3678869.3685683}
\acmISBN{979-8-4007-1113-8/24/09}

\AtBeginDocument{%
	}
		
\newcommand{\toolname}{\textsc{Gamekins IntelliJ plugin}\xspace}
\newcommand{\gamekins}{\textsc{\mbox{Gamekins}}\xspace}
\newcommand{\Jenkins}{\textsc{Jenkins}\xspace}

\newcommand{\University}{University of Passau}

\begin{document}
	
	\title{Engaging Developers in Exploratory Unit Testing \\through Gamification}

	\author{Philipp Straubinger}
	\affiliation{%
		\institution{University of Passau}
		\city{Passau}
		\country{Germany}
	}
	
	\author{Gordon Fraser}
	\affiliation{%
		\institution{University of Passau}
		\city{Passau}
		\country{Germany}
	}
	
	\renewcommand{\shortauthors}{Straubinger et al.}

	\begin{abstract}
          Exploratory testing, known for its flexibility and ability to uncover unexpected issues, often faces challenges in maintaining systematic coverage and producing reproducible results. To address these challenges, we investigate whether gamification of testing directly in the Integrated Development Environment (IDE) can guide exploratory testing. We therefore show challenges and quests generated by the \gamekins gamification system to make testing more engaging and seamlessly blend it with regular coding tasks. In a 60-minute experiment, we evaluated \gamekins' impact on test suite quality and bug detection. The results show that participants actively interacted with the tool, achieving nearly 90~\% line coverage and detecting 11 out of 14 bugs. Additionally, participants reported enjoying the experience, indicating that gamification can enhance developer participation in testing and improve software quality.
	\end{abstract}

	\ccsdesc[500]{Software and its engineering~Software testing and debugging}
	\ccsdesc[500]{Software and its engineering~Integrated and visual development environments}
	
	\keywords{Gamification, IDE, IntelliJ, Software Testing, Exploratory Testing}
	
	\maketitle
	
	\section{Introduction}
	Exploratory testing, a dynamic approach that prioritizes exploration and discovery, has become a common approach in software testing~\cite{DBLP:journals/sigsoft/Neri23}. Unlike traditional scripted testing, exploratory testing allows testers to design and execute tests on the fly, adapting to new information and discoveries as they go. This flexibility is crucial in identifying unexpected issues and understanding complex software behaviors that pre-defined test cases might overlook~\cite{DBLP:journals/ese/AfzalGITAB15}. However, despite its advantages, incorporating exploratory testing into established development workflows can be challenging. Testers often face difficulties in maintaining systematic coverage and providing reproducible results, and the approach can be perceived as less structured and rigorous compared to automated methods~\cite{DBLP:journals/ese/ItkonenM14}.

Developers, typically more focused on code creation than testing, find manual and therefore exploratory testing particularly unappealing~\cite{DBLP:conf/issre/StraubingerF23}. This lack of engagement can lead to insufficient testing, missed bugs, and lower overall software quality. Therefore, finding ways to make exploratory testing more attractive and engaging to developers is essential. This is where gamification can play a crucial role. Gamification involves incorporating game-like elements such as points, levels, challenges, and rewards into non-game activities to boost motivation and engagement~\cite{DBLP:conf/mindtrek/DeterdingDKN11}. In the context of software development, gamification can transform the testing process into a more enjoyable and stimulating activity, encouraging developers to participate more actively~\cite{DBLP:journals/csur/FulciniCAT23}.

To address the challenges of exploratory testing with gamification, we propose using \gamekins~\cite{DBLP:conf/icse/StraubingerF22}, which integrates gamified unit testing directly into the IDE or Continuous Integration (CI). Our goal is to seamlessly blend testing activities with developers' regular coding tasks, thereby reducing the friction of switching contexts. In our experiments, we, therefore, use the \toolname~\cite{DBLP:conf/icse/StraubingerFraser24}. \gamekins generates challenges and quests based on the codebase, motivating developers to write tests by rewarding their progress and achievements. The \toolname integrates exploratory testing at the unit level into the IDE by generating challenges based on an unfamiliar codebase. This helps reduce biases that arise from developers testing only their own code rather than collaborating with a testing expert~\cite{DBLP:conf/icimtech/AndriadiSGA23}. Since \gamekins focuses on unit testing, exploratory testing with it involves writing new unit tests to solve challenges, rather than manually testing the program. These new tests can later enhance the existing test suite.

The contributions of this paper are as follows:
\begin{itemize}
	\item We propose incorporating gamification into the IDE to facilitate exploratory unit testing.
	\item We introduce \gamekins as a tool to be used for exploratory unit testing in the IDE.
	\item We empirically evaluate the effects of \gamekins on the resulting test suites through a 60-minute experiment.
	\item We assess the effectiveness of these test suites in finding real-world bugs using a Defects4J dataset.
\end{itemize}

The study results show that participants actively engaged in exploratory unit testing using \gamekins. The resulting test suite achieved nearly 90~\% line coverage, over 70~\% mutation score, and detected 11 out of 14 bugs, demonstrating that exploratory unit testing can significantly enhance test suites. Feedback from participants indicates that while they enjoyed using \gamekins, there is room for further improvement.

	\section{Background}
	
\subsection{Gamification of Software Testing}

Testing is often perceived as tedious, stressful, uncreative, or unappreciated, which contributes to developers' reluctance to write tests and use test automation~\cite{DBLP:journals/corr/WaychalC16, DBLP:journals/software/WeyukerOBP00, kapur2017release, DBLP:conf/esem/SantosMCSCS17, DBLP:journals/infsof/DeakSS16}. Motivation, essential for developer productivity, can be intrinsic (an inner drive to engage in an activity) or extrinsic (external incentives related to the task)~\cite{DBLP:conf/esem/FrancaSS14, francca2014theory, 8370133}. Gamification, which incorporates game design elements into non-game contexts, provides extrinsic motivation~\cite{DBLP:conf/mindtrek/DeterdingDKN11}. Common gamification elements include points, badges, leaderboards, challenges, and achievements~\cite{robson2015all, DBLP:journals/csur/FulciniCAT23}. It has been demonstrated to boost engagement and improve outcomes compared to non-gamified environments~\cite{DBLP:conf/sast/JesusFPF18, DBLP:journals/ese/StolSG22, DBLP:conf/icse/Straubinger024, DBLP:conf/icse/Straubinger024a}.

Gamification has been shown to enhance student motivation in learning software testing~\cite{DBLP:conf/sbqs/JesusPFS19, DBLP:conf/icsob/Yordanova19}. There have also been efforts to gamify aspects of testing for professional developers, such as test-to-code traceability~\cite{DBLP:conf/sera/Parizi16}, acceptance~\cite{DBLP:conf/icse/ScherrEH18} and unit testing with CI~\cite{DBLP:conf/icse/Straubinger024a} and Integrated Development Environments (IDE)~\cite{DBLP:conf/icse/Straubinger024} support.

\subsection{IDE Support for Testing}

IDEs are essential tools in a developer's workflow, combining a code editor, compiler, debugger, UI builder, and other tools into a single application. These environments are highly customizable through plugins to meet specific requirements~\cite{1463097}. Most IDEs support writing and executing tests with various testing frameworks, simplifying the process of testing code. They include execution engines, powerful debuggers, and numerous tools to enhance code quality, such as test generation~\cite{DBLP:conf/icst/ArcuriCF16}, code coverage~\cite{DBLP:journals/csur/ZhuHM97}, mutation testing~\cite{5487526}, and test smell detection~\cite{van2001refactoring}. Despite offering all the necessary tools for effective testing, modern IDEs lack sufficient incentives to motivate developers to utilize these features fully. Previous work built plugins to gamify unit~\cite{DBLP:conf/icse/Straubinger024, DBLP:conf/icse/StraubingerFraser24} and GUI~\cite{DBLP:journals/corr/abs-2403-09842} testing in the IDE and showed their effectiveness to motivate developers to engage in testing activities.

\subsection{Exploratory Testing}

One widely adopted testing activity is exploratory testing, where testers simultaneously learn about the software, design tests, and execute them without predefined scripts. This method relies on the tester's creativity, experience, and intuition to discover unexpected issues and adapt to changes quickly. Conducted in time-boxed sessions, exploratory testing provides rapid feedback and enhances tester engagement~\cite{DBLP:journals/jss/MartenssonSMB21}. While it offers flexibility and can uncover unique bugs, its effectiveness depends on the tester's skill, and it often results in less formal documentation and reproducibility~\cite{DBLP:conf/icse/SuLXHX0024}.

Exploratory testing is typically conducted via the GUI because it provides testers with a visual representation of the system, allowing them to intuitively begin their exploration~\cite{DBLP:journals/tse/CoppolaFATA24, DBLP:conf/icst/FulciniA22}. Efforts have also been made to gamify the learning of exploratory testing~\cite{DBLP:conf/fie/CostaO20} and to gamify the process on a code level~\cite{DBLP:conf/gamify/Ozturk22}. In this work, we focus on gamifying exploratory testing at the unit level.


	\section{Exploratory Testing with \gamekins}
	

A major issue with exploratory testing is that it relies heavily on the experience and intuition of testers, which many developers lack~\cite{DBLP:journals/jss/MartenssonSMB21}. To address this, we propose using the challenges generated by \gamekins to guide developers in exploratory testing. This approach eliminates the need for developers to have prior experience and intuition in exploratory testing, as \gamekins takes on this role. \gamekins generates challenges based on test gaps related to coverage and mutation testing, directing developers to relevant parts of the code to begin their exploratory testing. While \gamekins is originally intended to be used via CI systems, directly showing the challenges in the IDE can inform developers during their exploratory testing efforts.

\subsection{The \toolname}

The \toolname integrates all gamification elements offered by \gamekins directly into the IDE. After logging into \gamekins within IntelliJ, all information is fetched through a customized API and displayed in a dedicated tab. This tab contains various pages showcasing the different gamification features of \gamekins. For more detailed information about the \toolname, please refer to \cite{DBLP:conf/icse/StraubingerFraser24}.

\subsection{Challenges}

\gamekins offers a variety of challenges that are test and quality-oriented tasks for the developer to solve. There are seven different types of challenges used in this study\footnote{Detailed information can be found in prior work~\cite{DBLP:conf/icse/StraubingerF22,DBLP:conf/icse/Straubinger024a}}:

\begin{itemize}
	\item \textbf{Build Challenge}: This challenge shows the developer that the failed build on the CI has to be fixed.
	\item \textbf{Test Challenge}: This challenge tasks to write a new test.
	\item \textbf{Class Coverage Challenge}: This challenge tasks the developer to cover more lines in a specific class.
	\item \textbf{Method Coverage Challenge}: This challenge focuses on improving the coverage of a specific method.
	\item \textbf{Line Coverage Challenge}: This challenge assigns the developer an uncovered line that they have to cover.
	\item \textbf{Branch Coverage Challenge}: This challenge focuses on the improvement of branch coverage of a covered line.
	\item \textbf{Mutation Challenge}: This challenge tasks the developer to detect a mutant generated by PIT\footnote{\url{https://pitest.org/}} by writing a new test.
\end{itemize}

\begin{figure}
	\includegraphics[width=\linewidth]{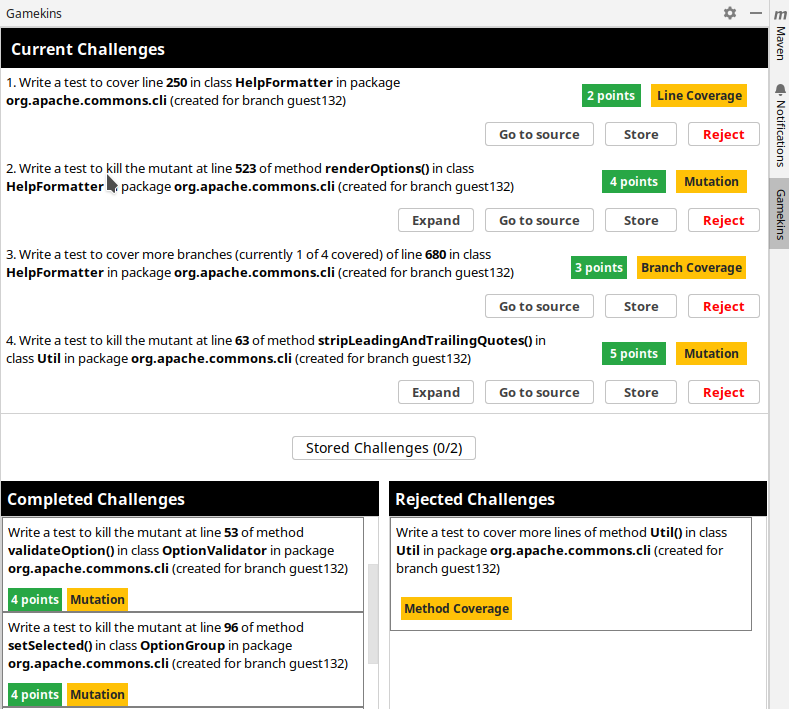}
	\vspace{-2em}
	\caption{Current, completed, and rejected challenges}
	\label{fig:challengesscreen}
\end{figure}

\begin{figure}
	\includegraphics[width=\linewidth]{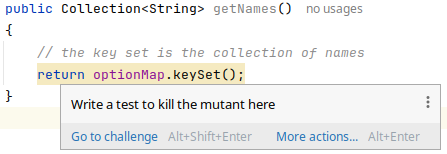}
	\vspace{-2em}
	\caption{Highlighted line of code with a tooltip giving information about the challenge}
	\label{fig:highlight}
\end{figure}

All challenges are displayed in the \toolname, each offering specific information and actions based on its type (\cref{fig:challengesscreen}). For example, Class and Method Coverage Challenges, as well as Line and Branch Coverage Challenges, feature buttons that allow users to navigate to and highlight the relevant code. Similarly, Mutation Challenges enable developers to highlight the original line of code and provide an expandable view to show the mutant.

Challenges are visually indicated in the source files with yellow highlights (\cref{fig:highlight}). When hovering over these highlighted sections, tooltips appear with a description and a link to the detailed view in the \toolname. This makes it easy for developers to identify and access information about the challenges directly from their code.

If a developer deems a challenge irrelevant, they can reject it with an explanation directly in the \toolname. Addressing challenges involves committing and pushing changes, which triggers \Jenkins to run the CI pipeline. After the pipeline execution, \gamekins analyses the results and displays notifications in IntelliJ for completed builds and any solved or new challenges.

\subsection{Quests}

\begin{figure}
	\includegraphics[width=\linewidth]{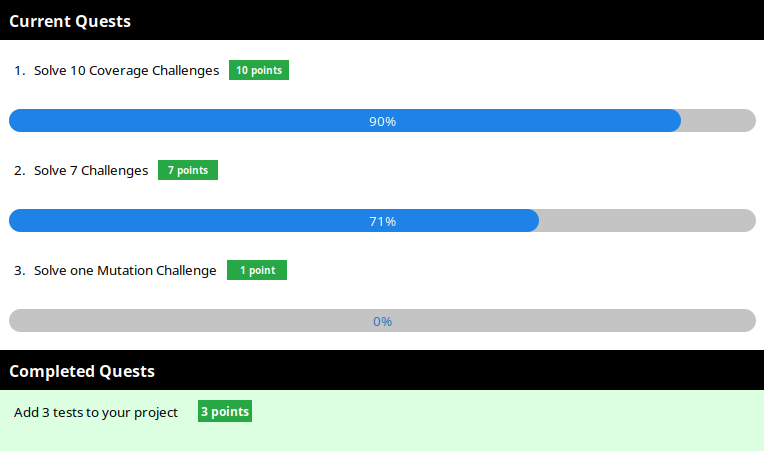}
	\vspace{-2em}
	\caption{Current and completed quests}
	\label{fig:questsscreen}
\end{figure}

Quests are tasks focused on testing and quality improvements, requiring multiple enhancements and interactions with gamification elements. In the \toolname, each quest is displayed on the quests page (\cref{fig:questsscreen}), accompanied by a description and a progress bar. This page shows both ongoing and completed quests. The progress bar, which dynamically updates with each relevant action, indicates the percentage of quest completion. The following types of quests are used in this study:
\begin{itemize}
	\item \textbf{Add Tests Quest}: This quest tasks the developer to add a specified number of tests to the existing test suite.
	\item \textbf{Cover Branches Quest}: This quest focuses on covering an additional specified number of branches with new tests.
	\item \textbf{Cover Lines Quest}: This quest focuses on covering an additional specified number of lines with new tests.
	\item \textbf{Solve Challenges Quest}: This quest tasks the developer to solve a specified number of new challenges of one type.
	\item \textbf{Solve Challenges Without Rejection Quest}: This quest focuses on solving a specified number of challenges regardless of their types without rejecting one in between.
\end{itemize}

\subsection{Achievements and Leaderboards}

Both achievements and leaderboards, while not the primary focus of this study, are integral parts of \gamekins. Developers earn rewards based on their testing accomplishments, which range from simple tasks like adding a test to more complex objectives such as achieving 100~\% coverage. Some achievements are hidden until completed and are tied to individual actions, such as adding new tests. There are two main types of achievements: individual, for completing a specific number of challenges, and project-level, for reaching certain project milestones. 

The \toolname includes all achievements from \gamekins, accessible through a dedicated tab displaying acquired and available achievements. Each achievement features an icon, title, description, and the date and time it was awarded. 

To motivate developers to solve challenges and complete quests, points are awarded based on task difficulty. These points contribute to user and team scores, which are prominently displayed on a leaderboard within the \toolname. The leaderboard shows rankings for individuals and teams, highlighting their scores, completed challenges, quests, achievements, and earned points. Additionally, users can personalize their appearance by choosing from 50 avatars available in \gamekins.

	\section{Experiment Setup}
	To evaluate whether exploratory testing with \gamekins in the IDE benefits the developers and the project, we conducted a controlled experiment and aim to answer the following research questions:
\begin{itemize}
	\item RQ1: How did the participants interact with \gamekins?
	\item RQ2: Does gamified exploratory testing lead to good test suites?
	\item RQ3: Can gamified exploratory testing find real-world bugs?
	\item RQ4: How did the participants perceive \gamekins?
\end{itemize}

\subsection{Experiment Task}

We aimed to use a real-world project to enhance the practical relevance of our experiment. The project needed to be comprehensible to participants yet not too simple. Additionally, to meet our exploratory testing objective, the project should not be entirely testable within our 60-minute time frame. The project also required a list of known fixed bugs, leading us to consider the Defects4J dataset\footnote{\url{https://github.com/rjust/defects4j}}. However, none of these projects met our criteria, while the Apache Commons CLI\footnote{\url{https://commons.apache.org/proper/commons-cli/}} project came closest to fulfilling our requirements. We simplified it by removing all deprecated classes, resulting in a project with six classes and 868 lines of code.

\subsection{Experiment Preparations}

The project was uploaded to a remote Git repository, where each participant has their own branch. These branches contain the task's source code, an example test, and a Jenkinsfile. A dedicated \Jenkins job is created for each branch, accessible only to the respective user. Each job is configured so that any push to the Git repository triggers a new build, while \gamekins generates challenges and quests specific to the current user.

In our lab setup, each participant has a designated computer. Each computer is equipped with a standard installation of IntelliJ Community Edition\footnote{\url{https://github.com/JetBrains/intellij-community}}, which includes the project and the corresponding branch opened. To streamline access without managing credentials, the project is cloned using a fine-grained access token. The \toolname is installed and linked to the user's \Jenkins job, displaying all relevant \gamekins features directly within IntelliJ. Additionally, an exit survey link is bookmarked in the browser for easy access after the experiment.

\subsection{Participants}

We designed a survey to recruit participants, including demographic questions, inquiries about programming experience in Java, familiarity with various Java testing tools, and five technical questions regarding JUnit to assess testing knowledge. Each technical question presented a small code example with four answer options\footnote{Detailed information is available in the artifacts}.

We advertised the survey among students in our Master's program and other PhD candidates at the University of Passau, resulting in 28 responses. As a minimum qualification, participants needed to answer at least three out of the five technical questions correctly to demonstrate expertise. Fifteen respondents met this criterion and were selected to participate.

Among the participants, eleven pursed a Master's degree at the time of the study, while four were PhD candidates in computer science. Some students also worked part-time in companies. The age of our participants ranged from 22 to 35, with one female participant. All but one participant had more than three years of overall experience, and nine out of 15 participants also had more than three years of experience specifically with JUnit testing.

\subsection{Experiment Procedure}

The experiment was conducted in four in-person sessions at the computer lab of the \University. Each session began with a 10-minute introduction to \gamekins and the task, which consisted of reating tests for the project by solving challenges through \gamekins. Participants could choose which challenges to solve and were allowed to reject any challenges. They were permitted to look up specifications on the internet, provided they did not use any form of Artificial Intelligence (AI).
After writing each test, participants were encouraged to commit and push their code to the remote Git repository. This action triggered \Jenkins to build the project and allowed \gamekins to check if challenges and quests were completed. While \Jenkins ran in the background, participants could proceed to the next challenge, as the plugin would notify them when the build was finished. After 60 minutes, participants submitted their current progress and completed an exit survey.

\subsection{Experiment Analysis}

The analysis of the experiment involves comparing the results obtained from the participating students and running the test suites against the Defects4J bug dataset.

\subsubsection{RQ1: How did the participants interact with \gamekins?}

The data collected by \gamekins, including current, completed and rejected challenges and quests, are stored in each user's configuration files. This data is easily extractable for further evaluation. We analyze the differences in number of challenges, quests, runs, and scores between participants. Additionally, we examine the various types of challenges they solved and investigate the reasons behind their rejection of certain challenges. This analysis helps us identify any difficulties users encountered while using \gamekins.

\subsubsection{RQ2: Does gamified exploratory testing lead to good test suites?}

In this research question, we examine and compare (1) the number of tests, (2) line coverage, (3) branch coverage, and (4) mutation scores among the participants. We measure line and branch coverage of their implementations using JaCoCo\footnote{\url{https://www.jacoco.org/jacoco/}}, and determine the mutation scores with PIT. Coverage is a common metric for assessing how thoroughly a project's source code is exercised~\cite{ivankovic2019code}, while mutation analysis helps identify test gaps within the covered code~\cite{DBLP:journals/tse/PetrovicIFJ22}. Finally, we merge all individual tests into a single comprehensive test suite and measure the same metrics as we did for the individual test suites. This approach provides a clearer understanding of the overall effectiveness of exploratory testing with multiple testers since in real-world projects, developers do not maintain separate test suites but one test suite for the entire project.

\subsubsection{RQ3: Can gamified exploratory testing find real-world bugs?}

Bugs from the Defects4J dataset are artificially reintroduced into the six classes of our project: \texttt{HelpFormatter}, \texttt{Option}, \texttt{OptionGroup}, \texttt{OptionValidator}, \texttt{Options}, and \texttt{Util}. The participants' tests are then executed (1) individually, (2) as a single test suite per participant, and (3) collectively with all participants' tests run against each bug. The results are compared between participants to analyze whether gamified exploratory testing can uncover real-world bugs.

\subsubsection{RQ4: How did the participants perceive \gamekins?}

\begin{table}[t]
	\centering
	\caption{Survey questions \\ \tiny{with Single Choice as SC}}
	\label{tab:allquestions}
	\vspace{-1em}
	\resizebox{\columnwidth}{!}{
		\begin{tabular}{lp{7cm}l}
			\toprule
			ID & Question & Type     \\ \midrule
			\addlinespace[0.5em]
			\multicolumn{2}{l}{Questions in the category participant demographics} \\ \cmidrule(r){1-2}
			P1                                                 & Age                                                                                            & Free-text       \\
			P2                                                 & Gender                                                                                         & SC + free-text  \\
			P3                                                 & Occupation                                                                                     & Free-text       \\
			P4                                                 & Experience with Java                                                                           & SC              \\
			P5                                                 & Experience with JUnit                                                                          & SC              \\
			\addlinespace[0.5em]
			\multicolumn{2}{l}{Questions in the category \gamekins} \\ \cmidrule(r){1-2}
			G1                                                 & The target project was easy to understand.                                                     & Likert 5 points \\
			G2                                                 & It was easy to write tests for the target project.                                             & Likert 5 points \\
			G3                                                 & I have produced a good test suite.                                                             & Likert 5 points \\
			G4                                                 & The plugin was intuitive to use.                                             & Likert 5 points \\
			G5                                                 & The plugin was easy to use.                                                  & Likert 5 points \\
			G6                                                 & I always knew how to solve challenges.                                                         & Likert 5 points \\
			G7                                                 & I always knew how to solve quests.                                                             & Likert 5 points \\
			G8                                                 & The notifications showed me my progress.                                                       & Likert 5 points \\
			G9                                                 & Having a plugin in the IDE is better than a browser-based version of \gamekins. & Likert 5 points \\
			G10                                                & I liked this part of the plugin -- Challenges                                    & Likert 5 points \\
			G11                                                & I liked this part of the plugin -- Quests                                        & Likert 5 points \\
			G12                                                & I liked this part of the plugin -- Achievements                                  & Likert 5 points \\
			G13                                                & I liked this part of the plugin -- Leaderboard                                   & Likert 5 points \\
			G14                                                & I liked this part of the plugin -- Notifications                                 & Likert 5 points \\
			G15                                                & I liked this part of the plugin -- Code Highlighting                             & Likert 5 points \\
			G16                                                & Have you used \gamekins before?                                                                 & Yes/No          \\
			G17                                                & What did you like about the plugin?                                          & Free-text       \\
			G18                                                & What did you dislike about the plugin?                                       & Free-text       \\
			G19                                                & How can the \toolname be improved?                                                   & Free-text       \\ 
			\bottomrule
		\end{tabular}%
	}\vspace{-1em}
\end{table}

To address this research question, we asked participants to complete a survey consisting of 29 questions divided into two categories (\cref{tab:allquestions}): demographics and experience with \gamekins, similar to previous studies~\cite{DBLP:conf/icse/Straubinger024a,DBLP:conf/icse/Straubinger024}. The demographic questions covered gender, occupation, and experience with testing. The second category focused on their thoughts about the task and \gamekins. We present the survey responses using Likert plots and analyze the students’ free-text answers to gain insights into their perceptions of \gamekins.

\subsection{Threats to Validity}

\textit{Threats to internal validity} may arise from participants' varying levels of experience with testing, which could influence the results. This risk is reduced by providing a default setup for all participants.

\textit{Threats to external validity}, which affect generalizability, include the limited sample of 15 Master's and PhD students. This threat is reduced because some participants also work in industry. Additionally, using a simplified project may not reflect the complexity of real-world projects. This is addressed by selecting the Apache Commons CLI project, which, while simple, is a real-world project.

\textit{Threats to construct validity} arise from conducting the experiment in a controlled lab with provided setups, which may not accurately represent how developers work on real-world projects.

	\section{Results}
	
\subsection{RQ1: How did the participants interact with \gamekins?}

\begin{figure}
	\centering
	\begin{subfigure}[t]{0.49\linewidth}
		\centering
		\includegraphics[width=\textwidth]{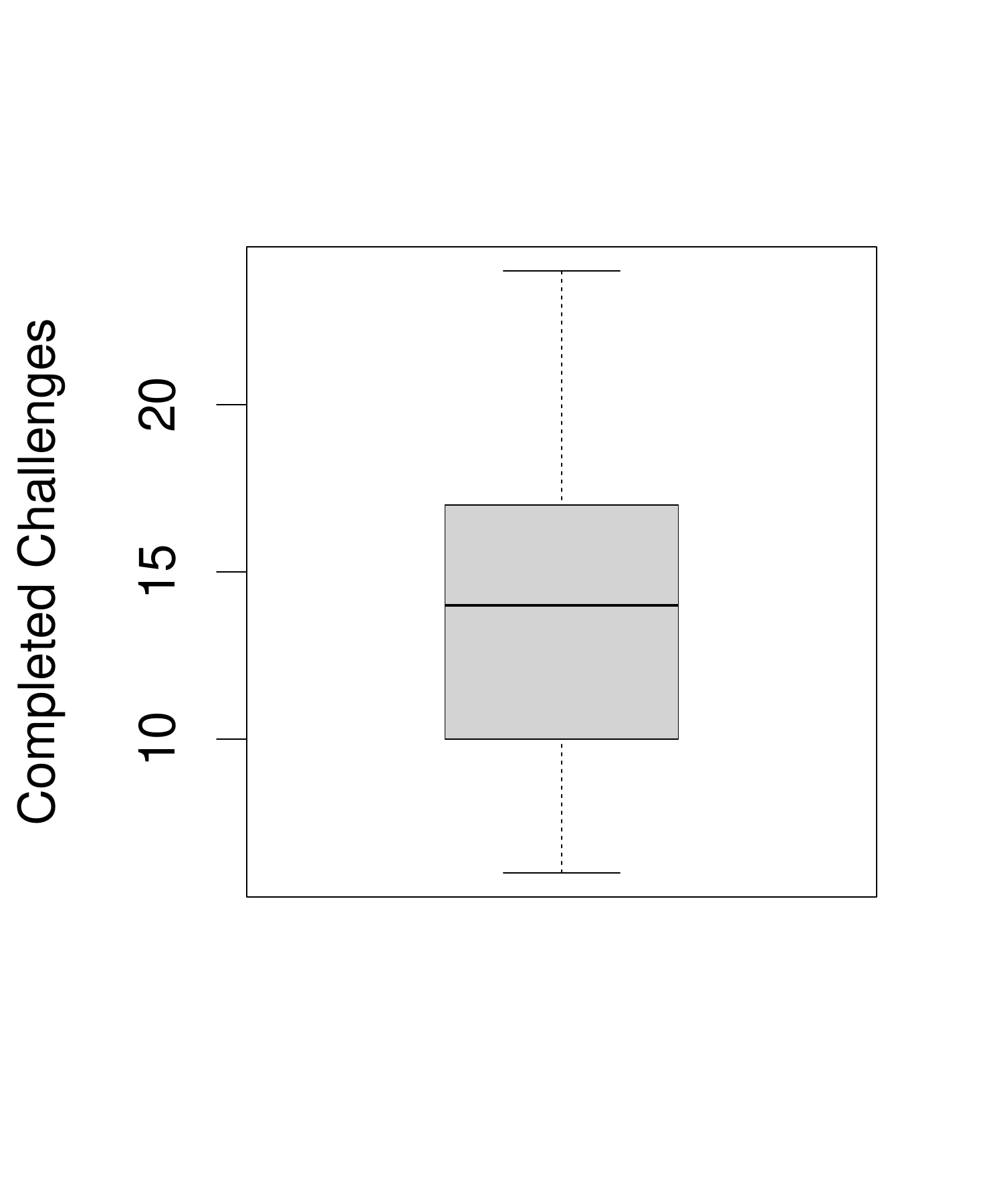}
		\vspace{-5em}
		\caption{Number of challenges}
		\label{fig:challenges}
	\end{subfigure}
	\hfill
	\begin{subfigure}[t]{0.49\linewidth}
		\centering
		\includegraphics[width=\textwidth]{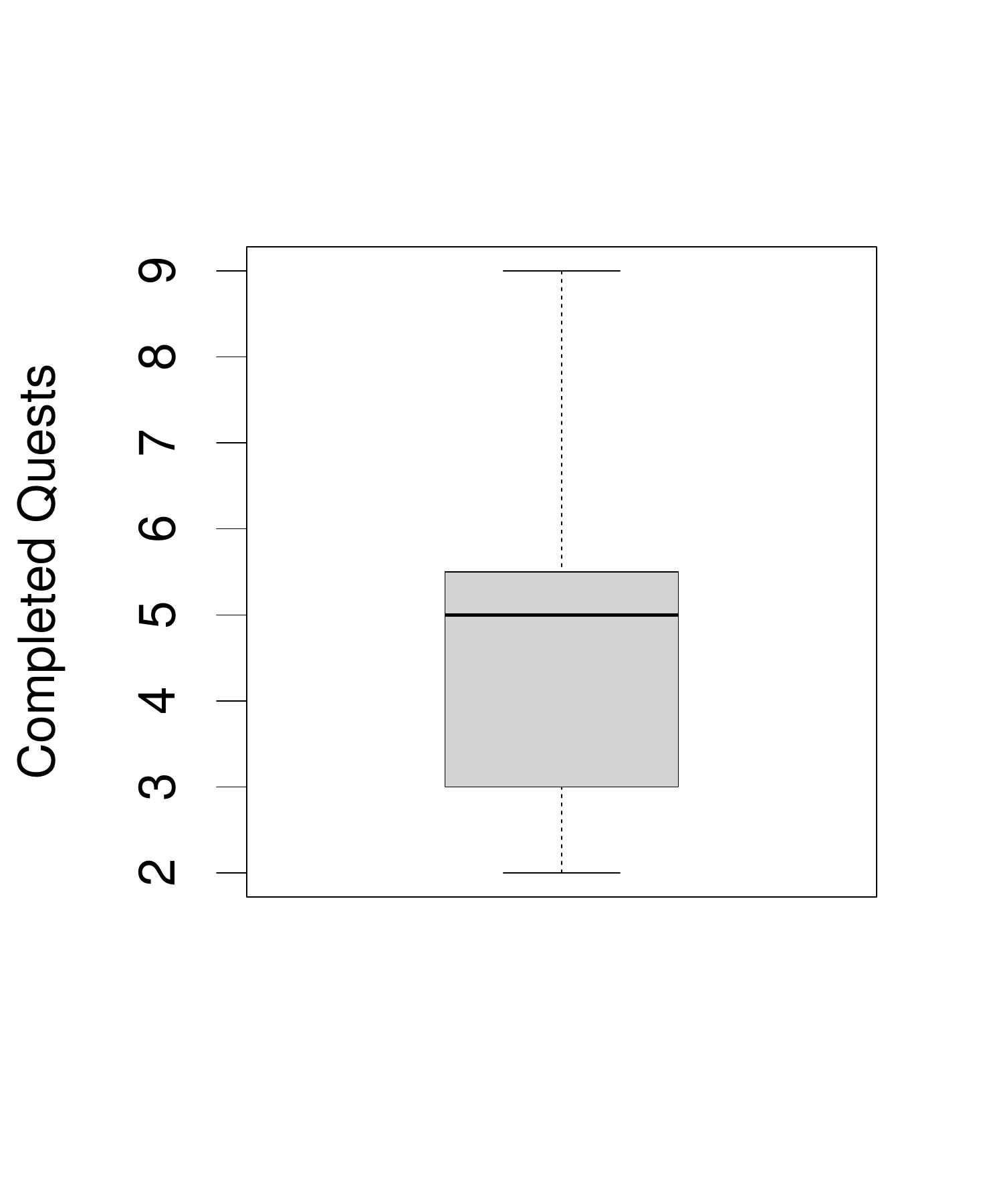}
		\vspace{-5em}
		\caption{Number of quests}
		\label{fig:quests}
	\end{subfigure}
	\hfill
	\vspace{-4em}
	\begin{subfigure}[t]{0.49\linewidth}
		\centering
		\includegraphics[width=\textwidth]{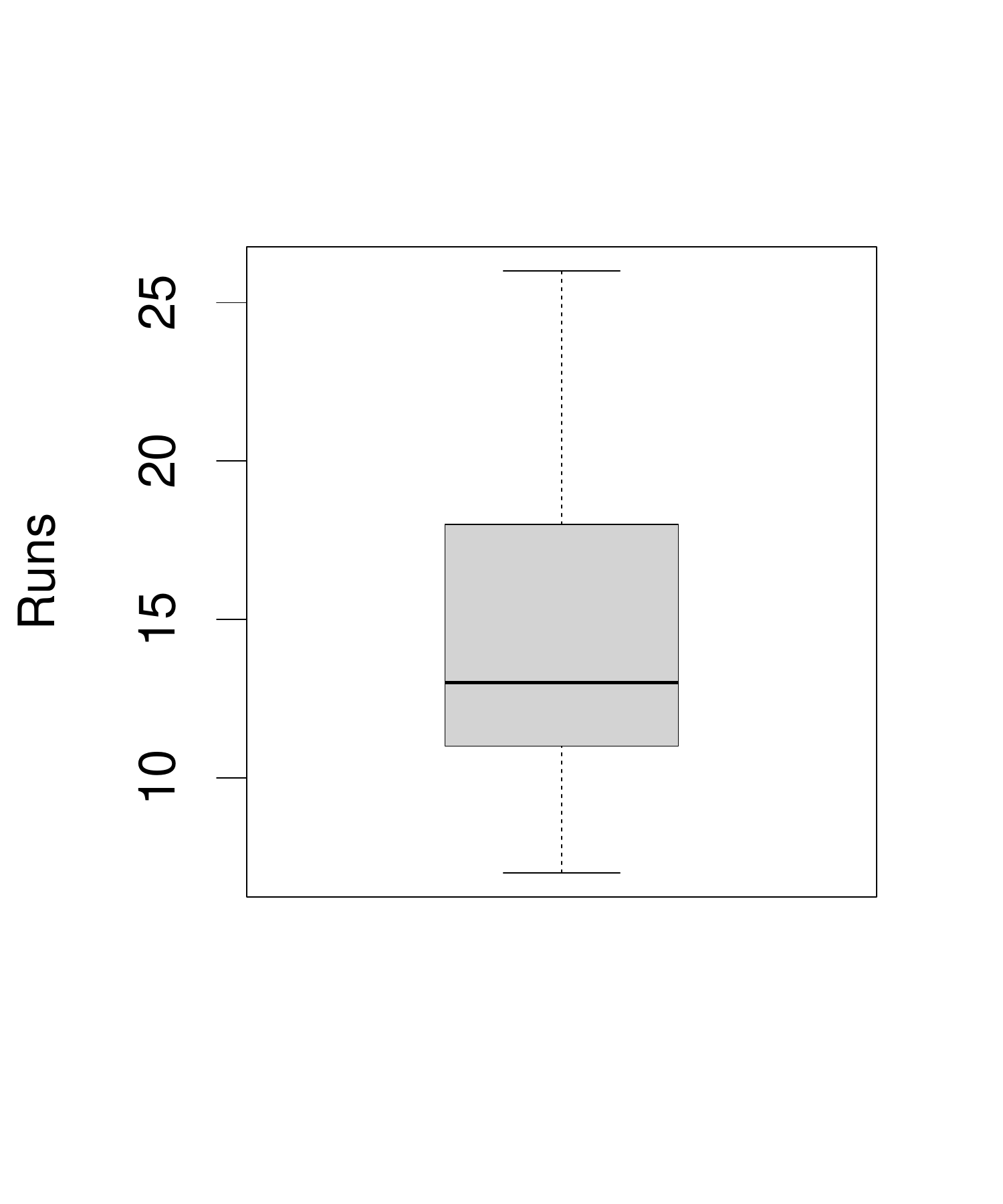}
		\vspace{-5em}
		\caption{Number of runs}
		\label{fig:runs}
	\end{subfigure}
	\hfill
	\begin{subfigure}[t]{0.49\linewidth}
		\centering
		\includegraphics[width=\textwidth]{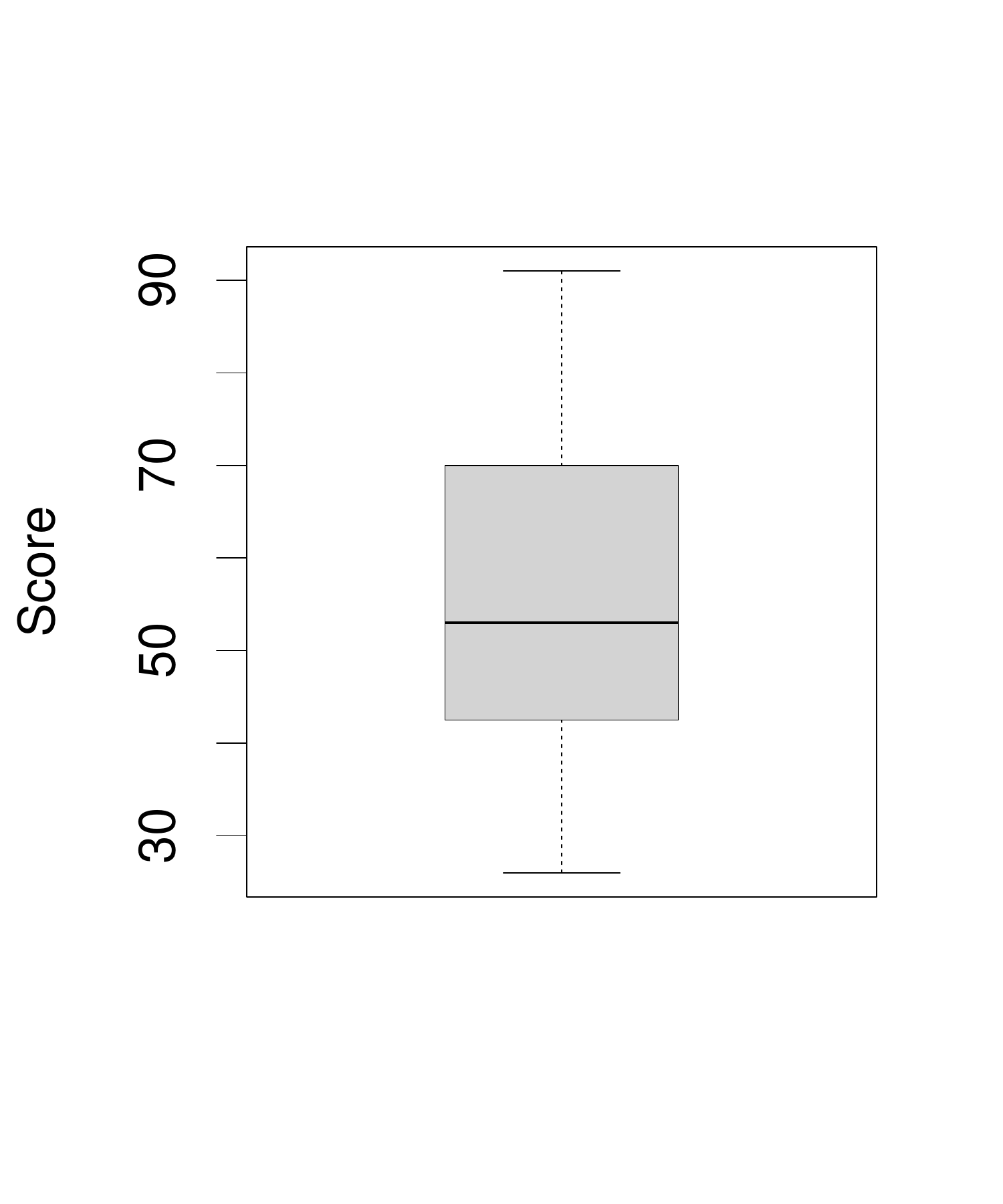}
		\vspace{-5em}
		\caption{Score in \gamekins}
		\label{fig:score}
	\end{subfigure}
	
	\vspace{-2em}
	\caption{Statistics on the use of \gamekins}
	\label{fig:statsplugin}
\end{figure}

The participants completed a total of 204 challenges and 71 quests, averaging 13.6 challenges (\cref{fig:challenges}) and 4.7 quests (\cref{fig:quests}) per participant. They achieved a mean score of 56.5 (\cref{fig:score}), with one participant scoring a maximum of 91 and another achieving a minimum of 26. Throughout the experiment, each participant worked independently without knowledge of the points accumulated by others, fostering competition among those who aimed to be on top of the leaderboard promised to be shown after the experiment.

Participants ran their projects 234 times, averaging 14.6 runs (\cref{fig:runs}). Since they lacked access to \Jenkins and could not trigger builds themselves, each run corresponds to a commit. Only two participants received a Build Challenge, indicating that they mostly ran tests locally before pushing to the remote repository. Mutation Challenges were most frequently completed with 58 instances, closely followed by 56 Line Coverage Challenges. This preference aligns with the challenges typically generated by \gamekins.

Eight challenges were rejected by five participants, indicating that only one-third encountered a reason to reject a challenge. The primary reason cited was discrepancies between IntelliJ's coverage information and JaCoCo's data used by \gamekins, leading participants to believe a line was covered when JaCoCo indicated otherwise. The second most common reason was uncertainty about testing private methods, suggesting a potential need for more education on testing strategies for private code segments.

\summary{RQ1}{Participants interacted with \gamekins by independently completing various challenges and quests, driven by a competitive leaderboard. They mostly preferred Mutation and Line Coverage challenges, running tests locally before committing, and only occasionally rejected challenges.}

\subsection{RQ2: Does gamified exploratory testing lead to good test suites?} \label{sec:rq2}

\begin{figure}
	\centering
	\begin{subfigure}[t]{0.49\linewidth}
		\centering
		\includegraphics[width=\textwidth]{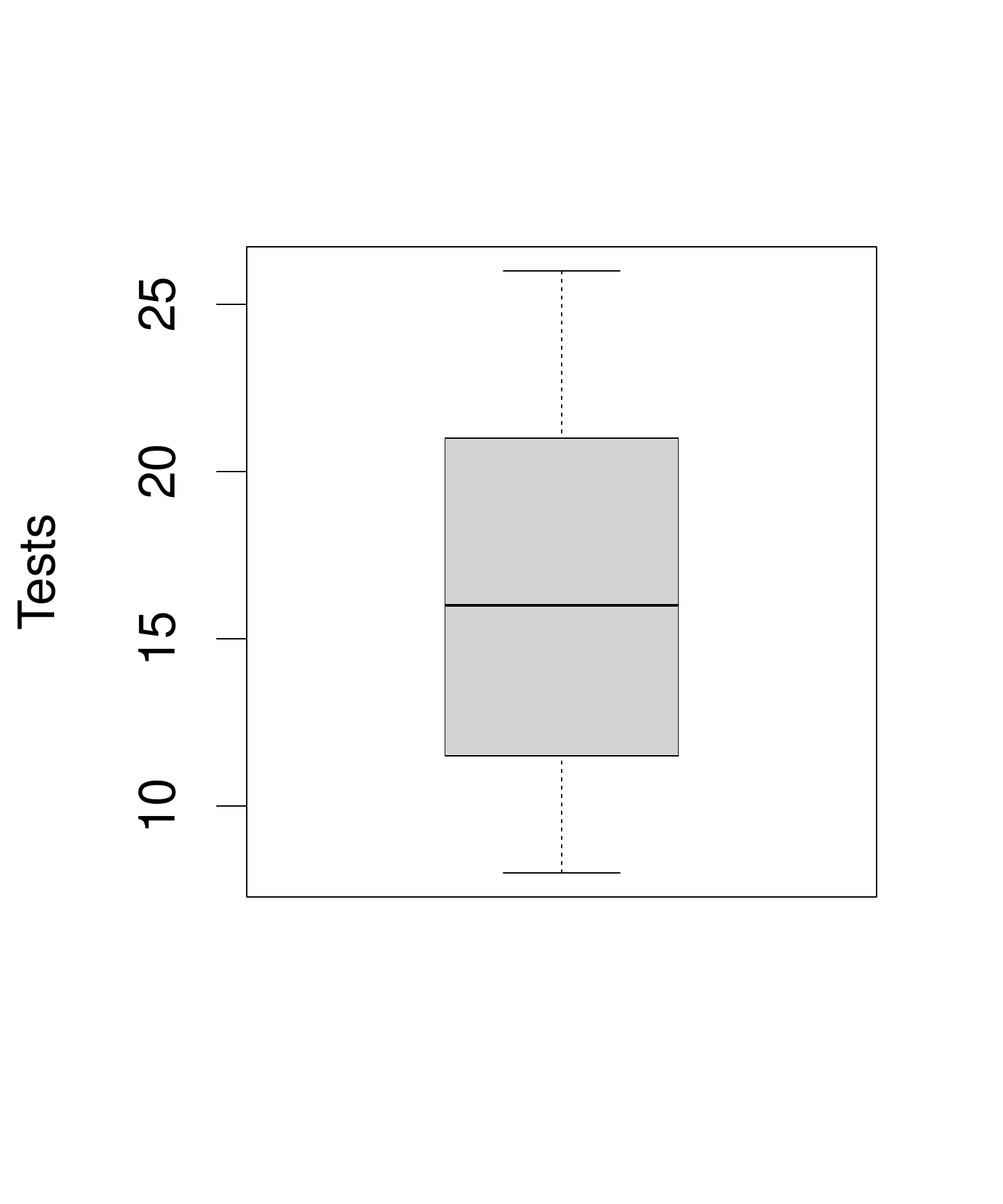}
		\vspace{-4em}
		\caption{Number of tests}
		\label{fig:tests}
	\end{subfigure}
	\hfill
	\begin{subfigure}[t]{0.49\linewidth}
		\centering
		\includegraphics[width=\textwidth]{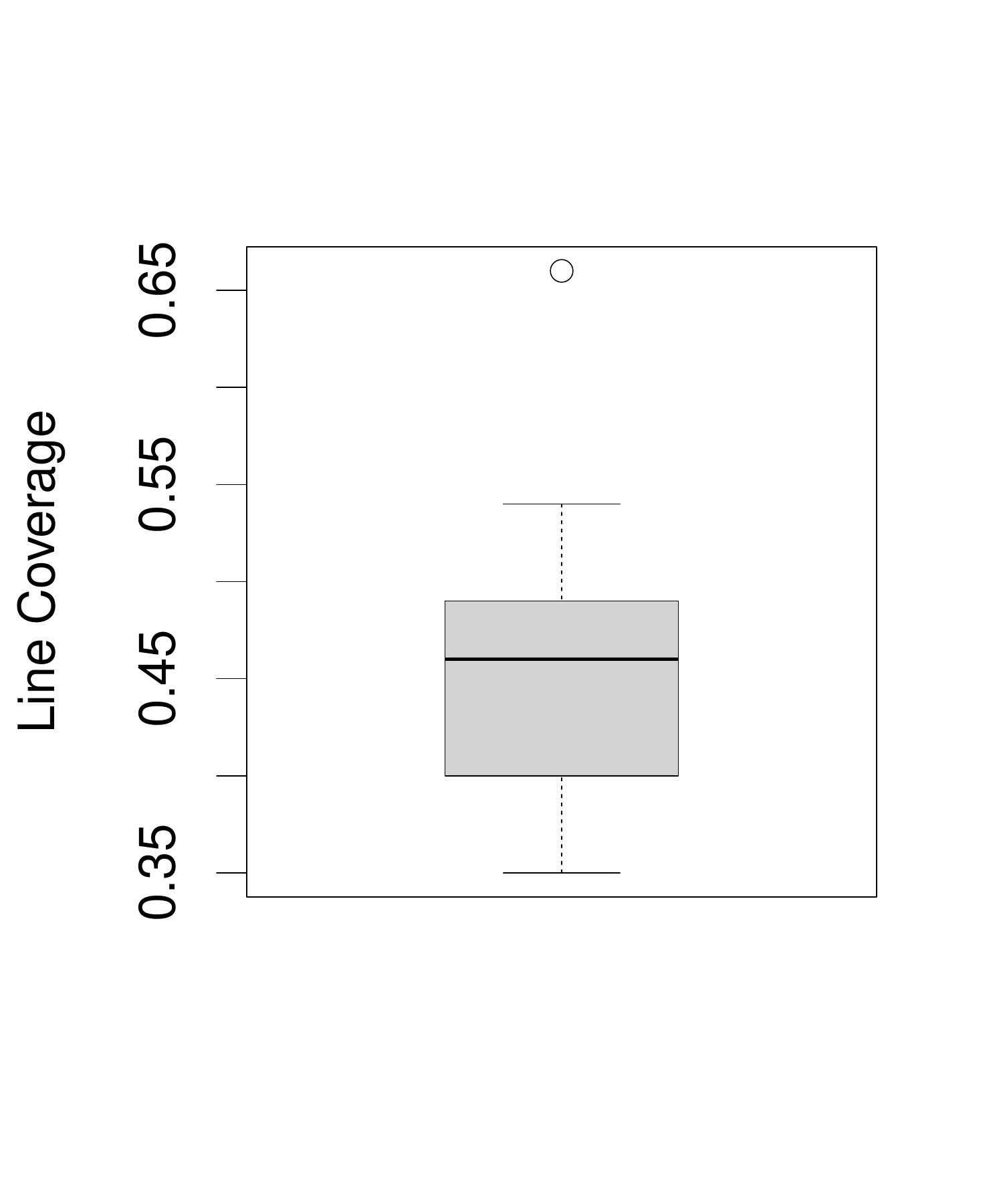}
		\vspace{-4em}
		\caption{Line coverage}
		\label{fig:linecoverage}
	\end{subfigure}
	\hfill
	\vspace{-4em}
	\begin{subfigure}[t]{0.49\linewidth}
		\centering
		\includegraphics[width=\textwidth]{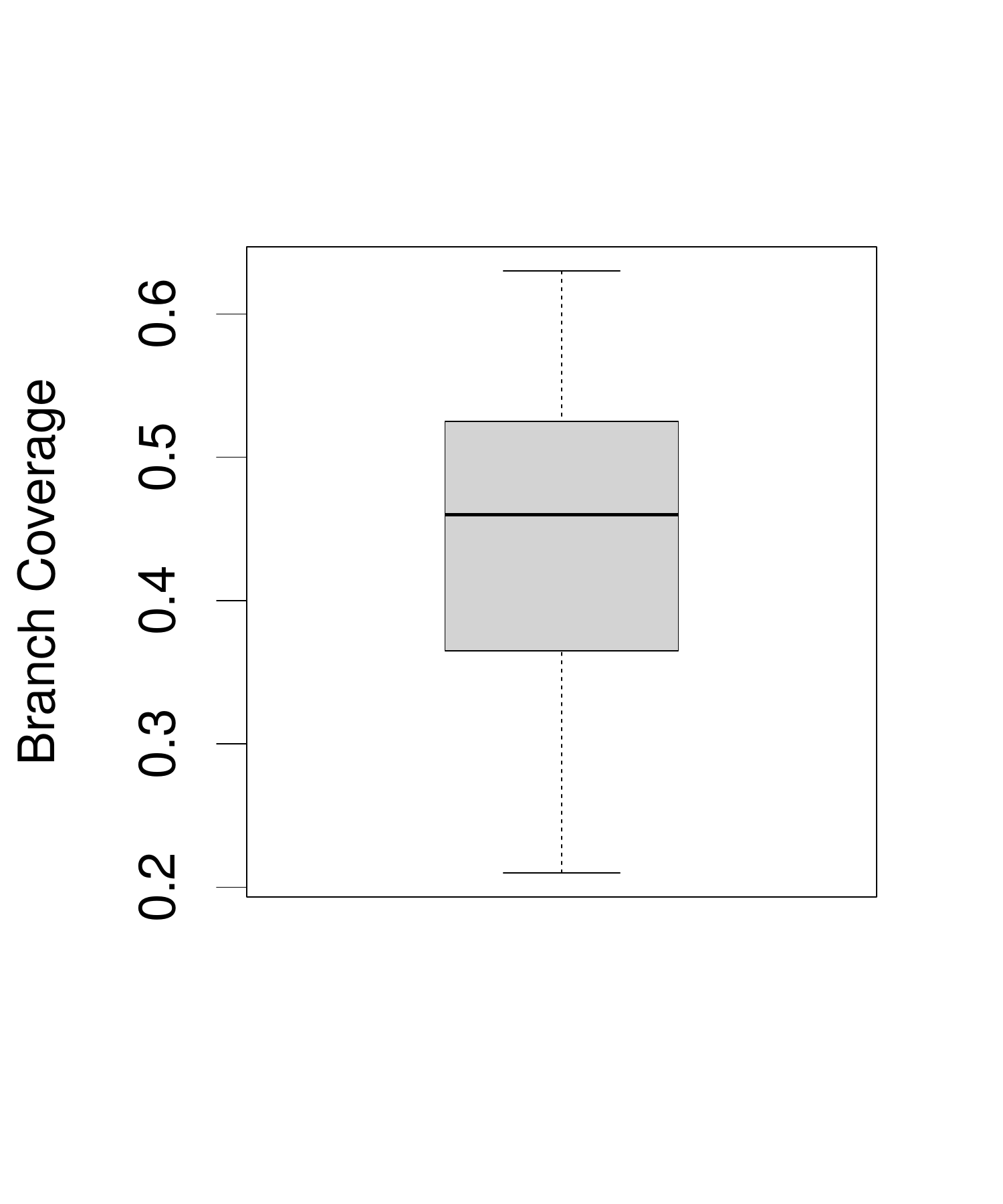}
		\vspace{-4em}
		\caption{Branch coverage}
		\label{fig:branchcoverage}
	\end{subfigure}
	\hfill
	\begin{subfigure}[t]{0.49\linewidth}
		\centering
		\includegraphics[width=\textwidth]{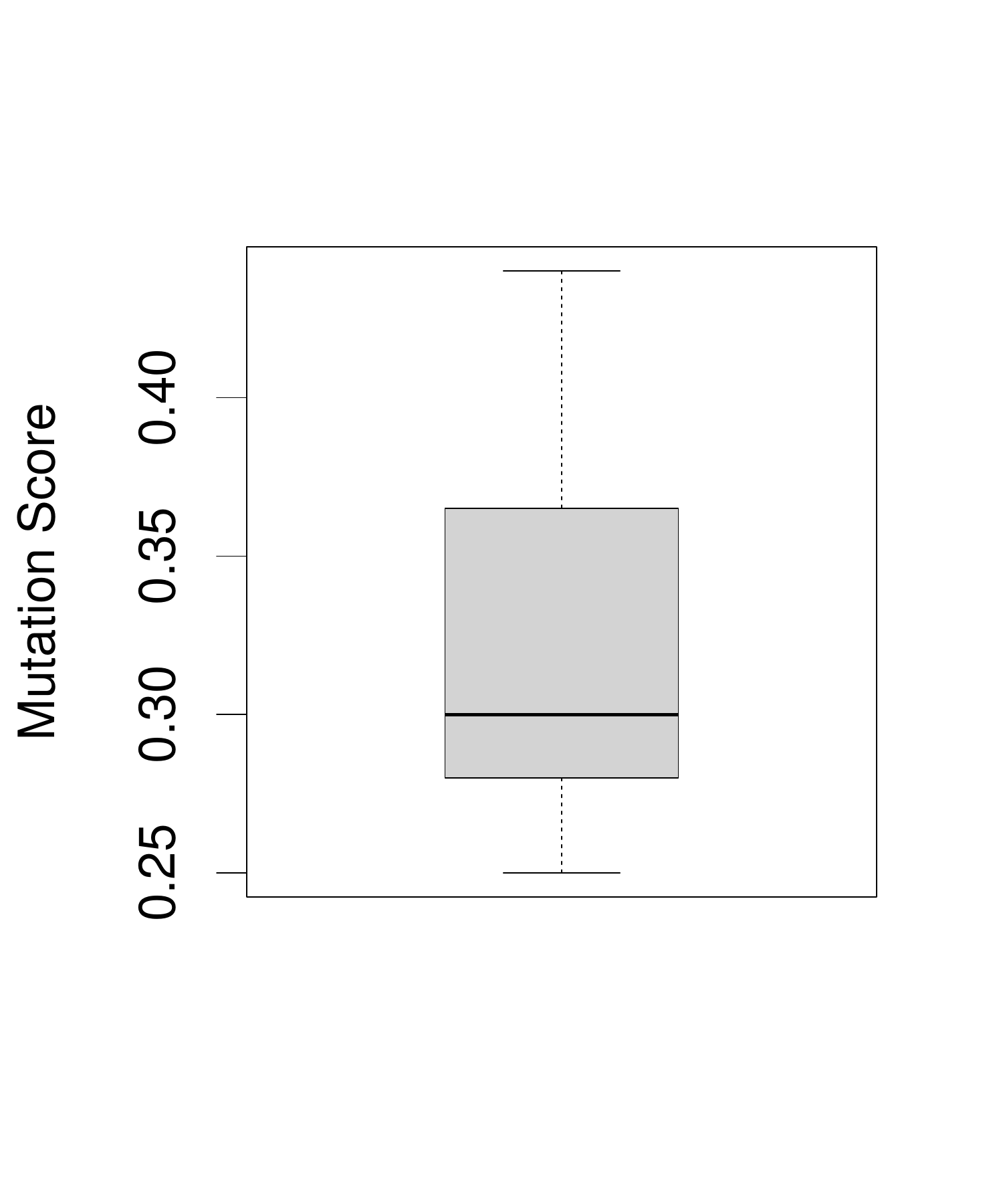}
		\vspace{-4em}
		\caption{Mutation score}
		\label{fig:mutationscore}
	\end{subfigure}
	
	\vspace{-1em}
	\caption{Statistics on the resulting test suites per participant}
	\label{fig:statssuite}
\end{figure}

On average, participants wrote 16.2 tests, with a maximum of 26 and a minimum of 8 (\cref{fig:tests}). They achieved an average line coverage of 46~\% (\cref{fig:linecoverage}) and 45~\% branch coverage (\cref{fig:branchcoverage}). These results indicate that \gamekins effectively balanced their efforts between achieving branch and line coverage. Regarding mutation score, participants achieved an average of 32~\% (\cref{fig:mutationscore}). Considering the experiment's limited 60-minute timeframe, participants managed to eliminate approximately one-third of the project's mutants.

Combining all participants' tests into a single suite of 243 tests, the suite achieved a total branch coverage of 68~\%. Since most branches are concentrated in the \texttt{HelpFormatter} and \texttt{Option} classes, other classes achieved nearly 100~\% branch coverage. In terms of line coverage, the suite reached 87~\%, indicating that almost all lines in the project were either targeted by \gamekins challenges or needed for solving them. The mutation score for the suite was 71~\%, showing a great improvement compared to individual participant scores. This collective approach demonstrates that the combined test suite is more robust than individual efforts.

\summary{RQ2}{Gamified exploratory testing led to good test suites, with participants achieving balanced line and branch coverage and a notable mutation score within a limited timeframe. The combined test suite from all participants showed significantly improved coverage and robustness compared to individual efforts.}
\begin{table}[t]
	\centering
	\caption{Bugs selected from the Defects4J dataset including their failed tests}
	\label{tab:bugs}
	\resizebox{\columnwidth}{!}{
		\begin{tabular}{llrrr}
			\toprule
			Bug & Class                 & Failed tests & Participants & Test targeting class \\ \midrule
			5   & Util                  & 11           & 11           & 10                   \\
			8   & HelpFormatter         & 6            & 3            & 6                    \\
			11  & HelpFormatter         & 0            & 0            & 0                    \\
			23  & HelpFormatter         & 2            & 2            & 2                    \\
			24  & HelpFormatter         & 2            & 2            & 2                    \\
			25  & HelpFormatter         & 2            & 2            & 2                    \\
			27  & OptionGroup           & 16           & 9            & 15                   \\
			29  & Util                  & 9            & 6            & 9                    \\
			31  & HelpFormatter, Option & 1            & 1            & 1                    \\
			32  & HelpFormatter         & 2            & 2            & 2                    \\
			33  & HelpFormatter         & 0            & 0            & 0                    \\
			34  & Option                & 8            & 6            & 3                    \\
			35  & Options               & 0            & 0            & 0                    \\
			36  & Options, OptionGroup  & 2            & 2            & 2                    \\
			\bottomrule
		\end{tabular}
	}
\end{table}

\vspace{0.9cm}
\begin{lstlisting}[language=Java,caption=The fixed code snippet of bug number 5,label=lst:5,escapechar=\%]
	public static String stripLeadingHyphens(final String str)
	{
		if (str == null)
		{
			return null;
		}
		if (str.startsWith("--"))
		{
			return str.substring(2);
		}
		if (str.startsWith("-"))
		{
			return str.substring(1);
		}
		
		return str;
	}
\end{lstlisting}

\begin{lstlisting}[language=Java,caption=The buggy code snippet of bug number 35,label=lst:buggy]
	public List<String> getMatchingOptions(String opt) {
		opt = Util.stripLeadingHyphens(opt);
		
		final List<String> matchingOpts = new ArrayList<>();
		
		for (final String longOpt : longOpts.keySet()) {
			if (longOpt.startsWith(opt)) {
				matchingOpts.add(longOpt);
			}
		}
		
		return matchingOpts;
	}
\end{lstlisting}

\begin{lstlisting}[language=Java,caption=The fixed code snippet of bug number 35,label=lst:fixed]
	public List<String> getMatchingOptions(String opt) {
		opt = Util.stripLeadingHyphens(opt);
		
		final List<String> matchingOpts = new ArrayList<>();
		
		// for a perfect match return the single option only
		if(longOpts.keySet().contains(opt)) {
			return Collections.singletonList(opt);
		}
		
		for (final String longOpt : longOpts.keySet()) {
			if (longOpt.startsWith(opt)) {
				matchingOpts.add(longOpt);
			}
		}
		
		return matchingOpts;
	}
\end{lstlisting}

\subsection{RQ3: Can gamified exploratory testing find real-world bugs?}

\begin{figure*}
	\centering
	\includegraphics[width=0.96\textwidth]{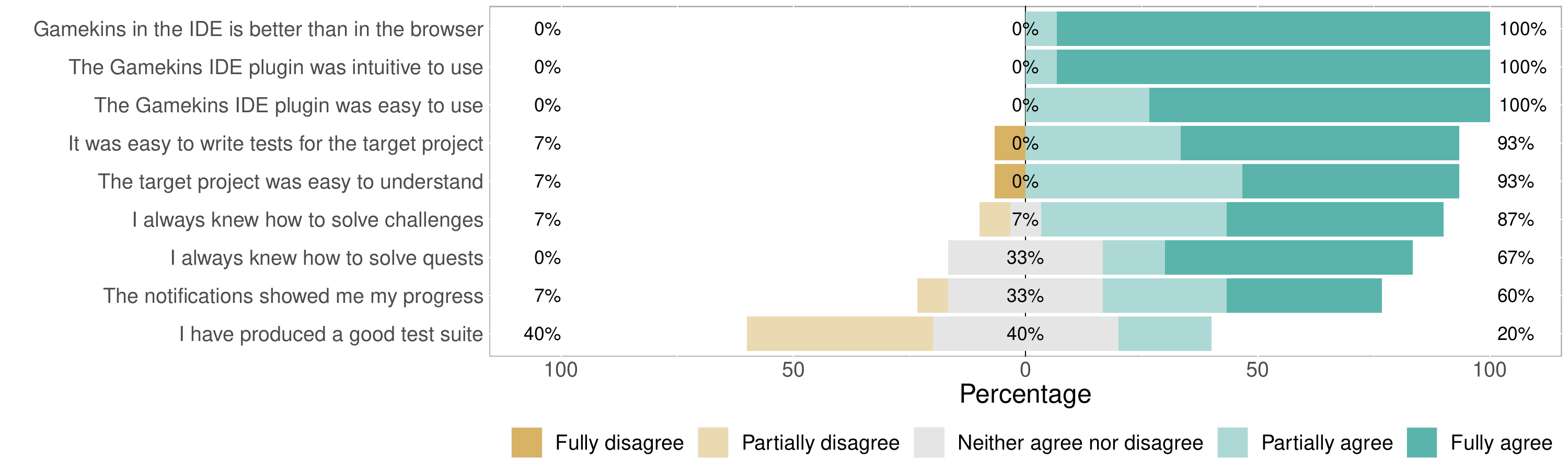}
	\vspace{-1em}
	\caption{Responses regarding the task and the usage of \gamekins}
	\label{fig:task}
\end{figure*}

\begin{figure*}
	\centering
	\includegraphics[width=0.73\textwidth]{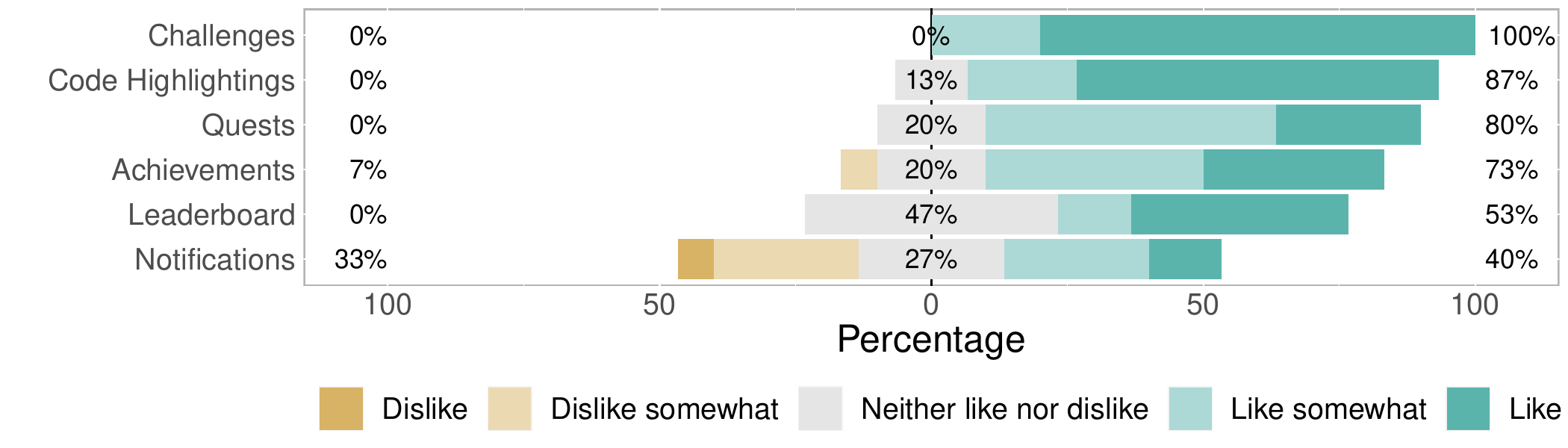}
	\vspace{-1em}
	\caption{Responses regarding the (gamification) elements of \gamekins}
	\label{fig:plugin}
\end{figure*}

The combined test suite created in \cref{sec:rq2} detected 11 out of 14 bugs, with the number of failed tests ranging from 1 to 16 out of the 243 tests in the suite (\cref{tab:bugs}). Most bugs were identified by just two failing tests, indicating a relatively low detection rate per bug. To better understand their significance, it is important to consider how many participants found each bug. Generally, the number of failed tests closely corresponds to the number of participants, with discrepancies typically occurring when a high number of tests fail.

It is noteworthy that most bugs were discovered by tests specifically targeting the classes where the bugs reside. Given that these bugs are concentrated in a small portion of the project, often involving only a few lines of modified code, it is likely that the tests revealing these bugs were created after \gamekins generated a challenge focusing on lines, branches, or mutants within them.

For instance, bug number 5 (\cref{lst:5}) was identified by most participants because the \texttt{Util} class is relatively compact, consisting of 29 lines in two methods. \gamekins prioritizes classes with low coverage, thus generating at least one challenge for \texttt{Util}, resulting in comprehensive coverage of the method containing the bug. The bug itself, involving a missing \texttt{null} check in the \texttt{getMatchingOptions} method, is straightforward to detect with a test.

However, three bugs went undetected by any tests. One such example is bug number 35 (\cref{lst:buggy}), where the function erroneously returns all options that start with the input string instead of only exact matches. For instance, if both \textit{package} and \textit{packageName} are valid options and the input is \textit{package}, the function returns both options instead of solely \textit{package}. This issue has been addressed by the maintainers, as illustrated in \cref{lst:fixed}, ensuring that only exact matches are returned. The remaining bugs involve incorrect if-conditions (bug number 11) and calls to different methods performing similar computations (bug number 33).

\summary{RQ3}{Gamified exploratory testing can effectively find real-world bugs, as the combined test suite detected most bugs, particularly in targeted classes. However, some of these bugs, requiring specific conditions while testing, remained undetected, indicating areas for improvement.}

\subsection{RQ4: How did the participants perceive \gamekins?}

In our study, participants included both those who had previously used \gamekins in the browser and those who were new to it. Everyone agreed that having a plugin within the IDE was preferable to switching to a browser for accessing features like the leaderboard, challenges, and quests (\cref{fig:task}). They also found \gamekins to be easy and intuitive to use, with a good understanding of how to tackle challenges and quests. The target project was generally straightforward for participants to comprehend and write tests for. However, they did not feel confident in producing a comprehensive test suite, likely due to the time constraint during the experiment.

Participants expressed overall positive sentiment toward almost all components of \gamekins (\cref{fig:plugin}), particularly appreciating the ability to navigate directly to challenges within the source code and the feature that highlights these challenges. Their feelings towards the leaderboard were mixed, leaning towards neutral and positive, possibly because they did not actively interact with it until seeing the final results at the end of the experiment.

One-third of the participants disliked the notifications from \gamekins, as detailed in their free-text responses at the end of the survey. They cited the high number of individual notifications after each build, including notifications for build results, solved challenges, quests, achievements, and newly generated challenges and quests. This frequent influx of notifications overwhelmed participants and diverted their focus from their tasks.

\summary{RQ4}{Participants perceived \gamekins positively, appreciating its integration within the IDE, ease of use, and intuitive interface for tackling challenges and quests. However, some found the frequent notifications overwhelming and distracting.}


	\section{Conclusions}
	In this study, we integrated gamified exploratory testing within the IDE using \gamekins, a tool designed to engage developers in the testing process through gamification. Our results showed that participants achieved high code coverage and successfully detected a significant number of bugs, indicating the effectiveness of the approach. Feedback from participants highlighted that the gamified elements made the testing process more engaging and enjoyable.

These findings suggest that gamification can significantly encourage active developer participation in exploratory testing, thereby enhancing software quality. By making the testing process more interactive and rewarding, developers are more motivated to create thorough and effective test suites.

For future work, we plan to enhance \gamekins with more sophisticated challenges, such as context-aware and scenario-based tasks, to further improve the quality of test suites. Additionally, we aim to refine the \toolname by reducing the number of notifications, as participants found the high frequency overwhelming, and by improving the plugin's GUI to make it more visually appealing and user-friendly.

	To support replications, all source code and experiment materials used in our study are available at:
	\begin{center}
		\url{https://doi.org/10.6084/m9.figshare.26402821}
	\end{center}
	
	\section*{Acknowledgements}
	\noindent This work is supported by the DFG under grant \mbox{FR 2955/2-1}, 
	``QuestWare: Gamifying the Quest for Software Tests''.

	\balance
	\bibliographystyle{ACM-Reference-Format}
	\bibliography{bib}
	\balance
	
\end{document}